\documentclass[11pt,a4paper]{article}
\usepackage{hyperref}
\usepackage[utf8]{inputenc}
\usepackage{latexsym,graphicx,amssymb,amsmath}
\usepackage{cite}
\usepackage{slashed}
\usepackage{bm}
\usepackage{float}
\usepackage{authblk}
\usepackage{abstract}
\usepackage{geometry}
\newcommand{\bea}{\begin{eqnarray}}
\newcommand{\eea}{\end{eqnarray}}
\newcommand{\be}{\begin{equation}}
\newcommand{\ee}{\end{equation}}
\begin{document}

\title{\vspace{-1cm}\textbf {Investigation of Pontryagin trace anomaly using Pauli-Villars regularization}}
\author{Chang-Yong Liu
\thanks{liuchangyong@nwsuaf.edu.cn}}
\affil{College of Science, Northwest A\&F University, Yangling, Shaanxi
712100, China}
\date{}
\maketitle

\begin{abstract}
In this paper, we investigate the Pontryagin trace anomaly for chiral fermions in a general curved background using Pauli-Villars regularization. We use both Feynman diagram method and Fujikawa's method to calculate the parity-odd contribution (Pontryagin term). Our result indicates that the trace anomaly of energy-momentum tensor for chiral fermions has Pontryagin term $P=\frac{i}{1536\pi^2}\epsilon_{\nu\sigma\kappa\lambda}R^{\nu\sigma}_{\ \ \rho\mu}R^{\rho\mu\kappa\lambda}$  which agrees with the work of Bonora et al \cite{Bonora:2014qla}.
\\
\begin{description}


\item[keywords] Trace anomaly, Pauli-Villars regularization, Chiral fermion

\end{description}
\end{abstract}



\section{Introduction}
Symmetries play an important role in our understanding of elementary particle physics. A symmetry of the classical action may be violated in the quantized version. This new feature of quantum theory was discovered in 1969 by Adler, Bell and Jackiw \cite{Adler:1969gk,Bell:1969ts} (Chiral anomaly) in the  solving the problem of neutral pion decay $\pi^0\rightarrow \gamma\gamma$. Gravitation is taken as a gauge theory also suffers from anomalies \cite{Alvarez-Gaume:1983ihn}. The gauges include the
general coordinate transformations (diffeomorphisms) and the conformal transformations (Weyl transformations). In this paper, we will study the trace or conformal anomaly.

We recall the definition of trace anomaly. The energy-momentum tensor in field theory is defined by
\bea
T_{\mu\nu}(x)=\frac{2}{\sqrt{|g|}}\frac{ \delta S}{\delta g^{\mu\nu}}.
\eea
 Consider the conformal transformation
\bea
\label{conformal1}
g_{\mu\nu}\rightarrow {\rm{e}}^{2 \sigma(x)}g_{\mu\nu}
\eea
for an infinitesimal value of the parameter $\sigma(x)$: $g_{\mu\nu}\rightarrow (1+2 \sigma(x) )g_{\mu\nu}$, we have
\bea
\label{delta}
\delta S=\frac{1}{2}\int d^4x\sqrt{|g|}T_{\mu\nu}\delta g^{\mu\nu}=-\int d^4x\ \sigma(x) \sqrt{|g|}T^{\mu}_{\mu}.
\eea
The invariance of $S$ under conformal transformation requires that trace of energy-momentum tensor is $T^{\mu}_{\mu}=0$. Generically, this classical traceless energy-momentum tensor is broken by quantum effects, that is
 \bea
 A=\langle T^{\mu}_{\mu}\rangle\neq 0.
 \eea
Here the $A$ is called trace or conformal anomaly \cite{Capper:1974ic,Capper:1975ig,Deser:1976yx,Bernard:1977pq,Brown:1977pq,Brown:1977sj,Christensen:1976vb,Adler:1976jx,
Duff:1977ay,Dowker:1976zf,Tsao:1977tj,
Christensen:1978gi,Vilenkin:1978wc,Wald:1978ce,Wald:1978pj,Christensen:1978md,Duff:1980qv}.

In four-dimension, the most general form of the trace anomaly was found to be given by \cite{Duff:1993wm,Bonora:1985cq,Bonora:1984ic}
\bea
\langle T^{\mu}_{\mu}\rangle=cF+aG+bR^2+b'\Box R+e \epsilon^{\alpha\beta\gamma\delta}R_{\alpha\beta\mu\nu}R^{\mu\nu}_{\gamma\delta},
\eea
where $F=R^{\alpha\beta\gamma\delta}R_{\alpha\beta\gamma\delta}-2 R^{\alpha\beta}R_{\alpha\beta}+\frac{1}{3}R^2$ is the square of the Weyl tensor and $G=R^{\alpha\beta\gamma\delta}R_{\alpha\beta\gamma\delta}-4 R^{\alpha\beta}R_{\alpha\beta}+R^2$ yields the Euler invariant. The coefficient $c$, $a$, $b$ and $b'$ are known at one-loop \cite{Parker:1978gh,Bertlmann:1996xk,Fujikawa:2004cx,Bastianelli:2006rx,Godazgar:2018boc,Frob:2019dgf}.
The last nontrivial term $\epsilon^{\alpha\beta\gamma\delta}R_{\alpha\beta\mu\nu}R^{\mu\nu}_{\gamma\delta}$ is the parity odd Pontryagin density which was first discussed by \cite{Nakayama:2012gu,Nakayama:2013oqa}. Later, the group of Bonora et al \cite{Bonora:2014qla} has
claimed that the parity odd Pontryagin density (CP-odd term) in the trace anomaly of Weyl fermions exists. The
coefficient of this term is purely imaginary signaling a violation of unitarity. These results were derived
both using a standard perturbative Feynman diagrams computation around Minkowski spacetime
 in dimensional regularization and the heat kernel method. For a spin $\frac{1}{2}$ right-handed spinor, the Weyl anomaly is connected with the Seeley-DeWitt coefficients \cite{DeWitt:1964mxt,Christensen:1978md,Vassilevich:2003xt,Bonora:2014qla}
\bea
\langle T^{\mu}_{\mu}\rangle=\frac{1}{180\times 16\pi^2}\left(-\frac{9}{2}F+\frac{11}{4}G+i\frac{15}{8} \epsilon^{\alpha\beta\gamma\delta}R_{\alpha\beta\mu\nu}R^{\mu\nu}_{\gamma\delta}\right).
\eea
Thus the Pontryagin term is
\bea
P=\frac{i}{1536\pi^2} \epsilon^{\alpha\beta\gamma\delta}R_{\alpha\beta\mu\nu}R^{\mu\nu}_{\gamma\delta}.
\eea

Our motivation for this work is that there are some different results for the Pontryagin term e.g. \cite{Bastianelli:2016nuf,Abdallah:2021eii,Abdallah:2022okt}. In this paper we are
going to recalculate the trace anomaly for a chiral fermion using Pauli-Villars regularization to try to figure it out. The paper is organized as follows. In Section 2, we use Feynman diagram method to calculate the parity-odd contribution. We regulate the divergence by introducing a set of Pauli-Villars fields. In Section 3, we apply the Fujikawa's method to evaluate the parity-odd term by Pauli-Villars regularization. We end with the conclusions. Some definitions and useful formulae are put in appendix.\\


\section{Feynman diagram method}
Following the works \cite{Bonora:2014qla,Bonora:2017gzz,Godazgar:2018boc,Abdallah:2021eii},
we consider the free Dirac fermion theory in 4d. The action is
\bea \label{action}
S=\int d^4x \sqrt{|g|}\left[i\overline{\psi}\gamma^{\mu}\left(\partial_{\mu}+\frac{1}{2}\omega_{\mu}\right)\psi-m\overline{\psi}\psi\right]=\int d^4x \sqrt{|g|}\left[i\overline{\psi}\gamma^{\mu}\nabla_{\mu}\psi-m\overline{\psi}\psi\right]
\eea
where the $\gamma^{\mu}$ is $\gamma^{\mu}=e^{\mu}_a \gamma^a$ ($\mu,\nu,\ldots$ are world indices, $a,b,\ldots$ are flat indices) and the $e^{\mu}_a$ is the inverse vierbein. The $\nabla_{\mu}=\partial_{\mu}+\frac{1}{2}\omega_{\mu}$ is the covariant derivative and $\omega_{\mu}$
is the spin connection:
\bea
\omega_{\mu}=\omega_{\mu}^{ab}\Sigma_{ab},
\eea
where $\Sigma_{ab}=\frac{1}{4}[\gamma_a,\gamma_b]$ are the Lorentz generators and the spin connection is
\bea
\omega_{\mu ab}=e_{\nu a}(\partial_{\mu} e^{\nu}_{b}+e^{\sigma}_{b} \Gamma_{\sigma\ \mu}^{\ \nu}).
\eea
The Levi-Civita connection $\Gamma_{\sigma\ \mu}^{\ \nu}$ is
\bea
\Gamma_{\sigma\ \mu}^{\ \nu}=\frac{1}{2}g^{\nu \rho}(\partial_{\sigma}g_{\mu \rho}+\partial_{\mu}g_{\sigma \rho}-\partial_{\rho}g_{\sigma \mu}).
\eea
We obtain the stress tensor by the following identities:
\bea
\delta \sqrt{|g|}&=&\frac{1}{2} \sqrt{|g|}g^{\mu\nu}\delta g_{\mu\nu},\nonumber\\
\delta \gamma^{\mu}&=&-\frac{1}{2}g^{\mu\nu}\gamma^{\rho} \delta g_{\rho\nu},\nonumber\\
\delta (\omega_{\mu\rho\sigma}\gamma^{\rho \sigma})&=&\gamma^{\rho \sigma}\nabla_{\sigma} \delta g_{\rho \mu}.
\eea
The action (\ref{action}) is
invariant under a local Lorentz transformation. From the action (\ref{action}), we can compute the stress tensor
\bea
\label{stresstensor}
\widehat{T}_{\mu\nu}(x)=\frac{2}{\sqrt{|g|}}\frac{ \delta S}{\delta g^{\mu\nu}}=-\frac{i}{4}\left(\overline{\psi}\gamma_{\mu}\overleftrightarrow{\nabla}_{\nu}\psi+(\mu\leftrightarrow\nu) \right)+g_{\mu\nu}
\left(\frac{i}{2}\overline{\psi}\gamma^{\lambda}\overleftrightarrow{\nabla}_{\lambda}\psi-m \overline{\psi}\psi \right).
\eea
Where we have introduced $\overleftrightarrow{\nabla}_{\mu}\equiv \nabla_{\mu}-\overleftarrow{\nabla}_{\mu}$. The $\overleftarrow{\nabla}_{\mu}$
acts according to $\overline{\psi}\overleftarrow{\nabla}_{\mu}=\partial_{\mu}\overline{\psi}-\frac{1}{2}\overline{\psi}\omega_{\mu}$. The ambiguity in the definition of the stress tensor leads to an ambiguity in the definition of
the trace anomaly. To eliminate these ambiguities, we use the definition of trace anomaly adopted by M.Duff \cite{Duff:1993wm}, which is given by the difference
\bea
g^{\mu \nu}\langle T_{\mu\nu}(x)\rangle-\langle g^{\mu\nu}T_{\mu\nu}(x)\rangle.
\eea
According to this definition, the second term of $\widehat{T}^{\mu\nu}(x)$ in (\ref{stresstensor}) drops out. Thus, we adopt the following stress tensor
\bea
T_{\mu\nu}(x)=-\frac{i}{4}\left(\overline{\psi}\gamma_{\mu}\overleftrightarrow{\nabla}_{\nu}\psi+(\mu\leftrightarrow\nu) \right).
\eea
Classically, the action (\ref{action}) is invariance under general coordinate transformation. It follows that

\bea
\partial_{\mu} T^{\mu \nu}(x)+\Gamma^{\mu}_{\mu\lambda}T^{\lambda\nu}(x)+\Gamma^{\nu}_{\mu\lambda}T^{\mu\lambda}=0.
\eea

\begin{figure}[H]
\center
\includegraphics[width=0.4\textwidth]{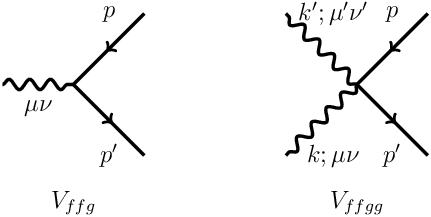}
\caption{\label{vertex}The two-fermion-one-graviton vertex ($V_{ffg}$) and two-fermion-two-graviton vertex ($V_{ffgg}$).  }
\end{figure}

We set $g_{\mu\nu}=\eta_{\mu\nu}+h_{\mu\nu}$, where $h_{\mu\nu}$ is a small perturbation around flat background. Using the following expansions
\bea
\label{expansion}
&&g^{\mu\nu}=\eta^{\mu\nu}-h^{\mu\nu}+(h^2)^{\mu\nu}+\ldots, \nonumber\\
&&\sqrt{|g|}=1+\frac{1}{2}({\rm{tr}}h)+\frac{1}{8}({\rm{tr}}h)^2-\frac{1}{4}h^{\mu\nu}h_{\mu\nu}+\ldots,\nonumber\\
&&e^{\mu}_{a}=\delta^{\mu}_{a}-\frac{1}{2}h^{\mu}_{a}+\frac{3}{8}(h^2)^{\mu}_{a}+\ldots,\nonumber\\
&&e^{a}_{\mu}=\delta^{a}_{\mu}+\frac{1}{2}h^{a}_{\mu}-\frac{1}{8}(h^2)^{a}_{\mu}+\ldots, \nonumber\\
&&\omega_{\mu ab}=\partial_{[b}h_{a]\mu}+\frac{1}{4}h^{\nu}_{[b|}\partial_{\mu}h_{|a]\nu}-\frac{1}{2}h^{\nu}_{[b|}(\partial_{\nu}h_{|a]\mu}-\partial_{a]}h_{\mu\nu})+\ldots.
\eea
and the relation
\bea
\{\gamma^a, \Sigma^{bc}\}=i\epsilon^{abcd}\gamma_d\gamma_5,
\eea
the action (\ref{action}) is expanded as
\bea
S&=& \int d^4 x \left[\frac{i}{2}(\delta^{\mu}_a-\frac{1}{2}h^{\mu}_a)\overline{\psi}\gamma^a\overleftrightarrow{\partial}_{\mu}\psi-m\overline{\psi}\psi+
\frac{1}{16}\epsilon^{\mu abc}\partial_{\mu}h_{a\lambda}h^{\lambda}_b\overline{\psi}\gamma_c\gamma_5\psi
\right]+\ldots \nonumber\\
&=&S^{(0)}+S^{(1)}+S^{(2)}+\ldots.
\eea
Where the $S^{(k)}$ is the order $k$ in the metric fluctuation $h_{\mu\nu}$, those are
\bea
S^{(0)}&=& \int d^4 x \left[\frac{i}{2}\overline{\psi}\gamma^{\mu}\overleftrightarrow{\partial}_{\mu}\psi-m\overline{\psi}\psi\right],\nonumber\\
S^{(1)}&=&-\frac{i}{4}\int d^4 x h^{\mu}_a\overline{\psi}\gamma^a\overleftrightarrow{\partial}_{\mu}\psi,\nonumber\\
S^{(2)}&=&\frac{1}{16}\int d^4 x  \epsilon^{\mu abc}\partial_{\mu}h_{a\lambda}h^{\lambda}_b\overline{\psi}\gamma_c\gamma_5\psi.
\eea
The Feynman rules can be read off directly from above action.
The fermion propagator, two-fermion-one-graviton vertex and two-fermion-two-graviton vertex (figure \ref{vertex}) are
\bea
\label{v}
P \quad &:& \quad \frac{i}{p\!\!\!/-m+i\epsilon}, \nonumber\\
V_{ffg} \quad &:& \quad -\frac{i}{8}\left[(p+p')_{\mu}\gamma_{\nu}+(p+p')_{\nu}\gamma_{\mu}\right], \nonumber\\
V_{ffgg} \quad &:& \quad \frac{1}{64}t_{\mu\nu\mu'\nu'\kappa\lambda}(k-k')^{\lambda}\gamma^{\kappa}\gamma^5, \nonumber\\
\eea
where the coefficient $t_{\mu\nu\mu'\nu'\kappa\lambda}$ is
\bea
t_{\mu\nu\mu'\nu'\kappa\lambda}=\eta_{\mu\mu'}\epsilon_{\nu\nu'\kappa\lambda}+\eta_{\nu\nu'}\epsilon_{\mu\mu'\kappa\lambda}
+\eta_{\mu\nu'}\epsilon_{\nu\mu'\kappa\lambda}+\eta_{\nu\mu'}\epsilon_{\mu\nu'\kappa\lambda}.
\eea

\par

To discuss the trace anomaly for chiral fermions, we define a chiral stress tensor
\bea
{T^{(R)}}_{\mu\nu}(x)=-\frac{i}{4}\left(\overline{\psi}\gamma_{\mu}\overleftrightarrow{\nabla}_{\nu}P_R\psi+(\mu\leftrightarrow\nu) \right).
\eea
Using the equation of motion for $\psi$, the trace of ${T^{(R)}}_{\mu\nu}(x)$ become
\bea
{\overline{T}^{(R)}}^{\mu}_{\ \mu}(x)=-m \overline{\psi}P_R\psi.
\eea
Then the trace anomaly of chiral stress tensor is
\bea
\label{traceanomaly}
A=\lim_{m\rightarrow 0}\langle {T^{(R)}}^{\mu}_{\ \mu}(x)+m \overline{\psi}P_R\psi\rangle=\lim_{m\rightarrow 0}\langle {\overline{T}^{(R)}}^{\mu}_{\ \mu}(x)\rangle.
\eea
Where the modified chiral stress tensor ${\overline{T}^{(R)}}_{\mu\nu}(x)$ is defined as
\bea
\label{modified}
{\overline{T}^{(R)}}_{\mu\nu}(x)=-\frac{i}{4}\left(\overline{\psi}\gamma_{\mu}\overleftrightarrow{\nabla}_{\nu}P_R\psi+(\mu\leftrightarrow\nu) \right)+\frac{m}{4}g_{\mu\nu} \overline{\psi}P_R\psi.
\eea
 We will directly use the expansion (\ref{expansion})
 together with Wick's theorem to evaluate the trace anomaly (\ref{traceanomaly}). The modified chiral stress tensor (\ref{modified}) also admits an expansion
 \bea
 {\overline{T}^{(R)}}_{\mu\nu}(x)={\overline{T}^{(R,0)}}_{\mu\nu}(x)+{\overline{T}^{(R,1)}}_{\mu\nu}(x)+\ldots,
 \eea
 where, to first order in $h_{\mu\nu}$,
 \bea
 \label{feynmandiagram2}
 {\overline{T}^{(R,0)}}_{\mu\nu}(x)&=&-\frac{i}{4}\left(\overline{\psi}\gamma_{\mu}\overleftrightarrow{\partial}_{\nu}P_R\psi+(\mu\leftrightarrow\nu) \right)+\frac{m}{4}\eta_{\mu\nu} \overline{\psi}P_R\psi, \nonumber\\
 {\overline{T}^{(R,1)}}_{\mu\nu}(x)&=&\frac{1}{2}\epsilon_{\mu}^{ abd}\overline{\psi}\partial_{[a}h_{b]\nu}\gamma_d\gamma_5P_R\psi+\frac{m}{4}h_{\mu\nu} \overline{\psi}P_R\psi.
 \eea

\begin{figure}[H]
\center
\includegraphics[width=0.3\textwidth]{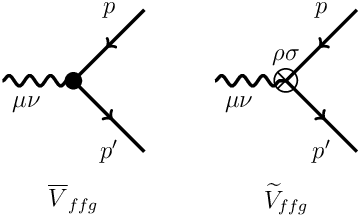}
\caption{\label{vertex1} Two new two-fermion-one-graviton vertex $\overline{V}_{ffg}$ and $\widetilde{V}_{ffg}$.  }
\end{figure}

 To use the intuitive Feynman diagram method, we introduce two new two-fermion-one-graviton vertices $\overline{V}_{ffg}$ and $\widetilde{V}_{ffg}$ (figure \ref{vertex1}), which come from the ${\overline{T}^{(R,0)}}_{\mu\nu}(x)$ and ${\overline{T}^{(R,1)}}_{\mu\nu}(x)$ separately.
From the expression (\ref{feynmandiagram2}), we obtain the two new vertices to be
\bea
&&\overline{V}_{ffg} \quad : \quad -\frac{i}{8}\left[(p+p')_{\mu}\gamma_{\nu}+(p+p')_{\nu}\gamma_{\mu}-m\eta_{\mu\nu}\right]\frac{1+\gamma^5}{2}, \nonumber\\
&&\widetilde{V}_{ffg} \quad : \quad -[\eta_{\sigma\nu}\epsilon_{\rho\mu ad}+\eta_{\sigma\mu}\epsilon_{\rho\nu ad}](p'-p)^a\gamma^d\frac{1+\gamma^5}{2}+\frac{m}{8}
(\eta_{\mu\rho}\eta_{\nu\sigma}+\eta_{\mu\sigma}\eta_{\nu\rho})\frac{1+\gamma^5}{2}.\nonumber
\eea
The expectation value of the modified chiral stress tensor trace ${\overline{T}^{(R)}}^{\sigma}_{\sigma}(x)$ in the metric perturbation is
\bea
\Big\langle{\overline{T}^{(R)}}^{\sigma}_{\sigma}(x)\Big\rangle&=&\Big\langle{\overline{T}^{(R)}}^{\sigma}_{\sigma}(x) {\rm{e}}^{i(S^{(1)}+S^{(2)}+\ldots)}\Big\rangle_0 \nonumber\\
&=&\Big\langle\left({\overline{T}^{(R,0)}}^{\sigma}_{\sigma}(x)+{\overline{T}^{(R,1)}}^{\sigma}_{\sigma}(x)+\ldots\right)\left(1+iS^{(1)}+(iS^{(2)}
-\frac{1}{2}S^{(1)}S^{(1)})+\ldots\right)\Big\rangle_0
\nonumber\\
&=&i\Big\langle{\overline{T}^{(R,0)}}^{\sigma}_{\sigma}(x)S^{(1)}\Big\rangle_0+i\Big\langle{\overline{T}^{(R,1)}}^{\sigma}_{\sigma}(x)S^{(1)}\Big\rangle_0
+i\Big\langle{\overline{T}^{(R,0)}}^{\sigma}_{\sigma}(x)S^{(2)}\Big\rangle_0 \nonumber\\
&&-\frac{1}{2}\Big\langle{\overline{T}^{(R,0)}}^{\sigma}_{\sigma}(x)S^{(1)}S^{(1)}\Big\rangle_0+\ldots
\eea
where the $\langle\cdots\rangle_0$ denotes expectation value in the free theory with action $S^{(0)}$. We have omitted the tadpole diagrams which have vanish contributions. The $\Big\langle{\overline{T}^{(R)}}^{\sigma}_{\sigma}(x)\Big\rangle$
can be separated into parity-odd term and parity-even term
\bea
\Big\langle{\overline{T}^{(R)}}^{\sigma}_{\sigma}(x)\Big\rangle=\Big\langle{\overline{T}^{(R)}}^{\sigma}_{\sigma}(x)\Big\rangle^{odd}
+\Big\langle{\overline{T}^{(R)}}^{\sigma}_{\sigma}(x)\Big\rangle^{even}.
\eea
In this paper, we only consider the parity-odd terms which contain the Levi-Civita symbol $\epsilon^{\mu\nu\rho\sigma}$ factor.

\begin{figure}[H]
\center
\includegraphics[width=0.4\textwidth]{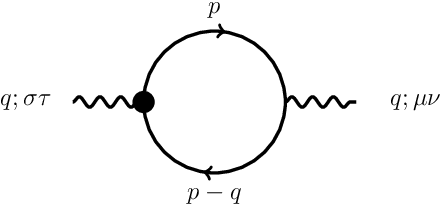}
\caption{\label{oneloop2} Feynman diagram corresponding to $i\Big\langle{\overline{T}^{(R,0)}}^{\sigma\tau}(x)S^{(1)}\Big\rangle_0$ term. }
\end{figure}

We first consider the expectation value of ${\overline{T}^{(R)}}^{\sigma}_{\sigma}(x)$ at $\mathcal{O}(h)$ order that is
\bea
\Big\langle{\overline{T}^{(R)}}^{\sigma}_{\sigma}(x)\Big\rangle\Big|_{\mathcal{O}(h)}=i\Big\langle{\overline{T}^{(R,0)}}^{\sigma}_{\sigma}(x)S^{(1)}\Big\rangle_0
=\int d^4y \int \frac{d^4 q}{(2\pi)^4}{\rm{e}}^{-iq\cdot(x-y)}T^{\sigma}_{\sigma\mu\nu}(q,m)h^{\mu\nu}(y).\nonumber
\eea
Using the Feynman rules (figure \ref{oneloop2}), the $T^{\sigma}_{\sigma\mu\nu}(q,m)$ is
\bea
\label{TToneloop2}
T^{\sigma}_{\sigma\mu\nu}(q,m)=\frac{i}{32}\int \frac{d^4 p}{(2\pi)^4}{\rm{tr}}\left[ \frac{1}{p\!\!\!/-m}\left((2p-q)_{\mu}\gamma_{\nu}+(\mu\leftrightarrow\nu)\right) \frac{1}{p\!\!\!/-q\!\!\!/-m}
(2p\!\!\!/-q\!\!\!/-2m)\frac{1+\gamma^5}{2}\right].\nonumber
\eea
\par
After calculating the integral, there have no parity-odd term in $T^{\sigma}_{\sigma\mu\nu}(q,m)$.

\begin{figure}[H]
\center
\includegraphics[width=0.4\textwidth]{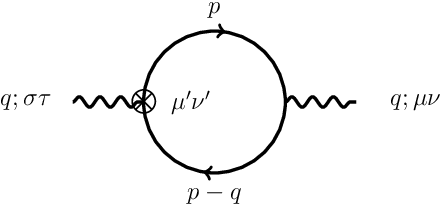}
\caption{\label{oneloop3} Feynman diagram corresponding to $i\Big\langle{\overline{T}^{(R,1)}}^{\sigma\tau}(x)S^{(1)}\Big\rangle_0$ term. }
\end{figure}

\begin{figure}[H]
\center
\includegraphics[width=0.4\textwidth]{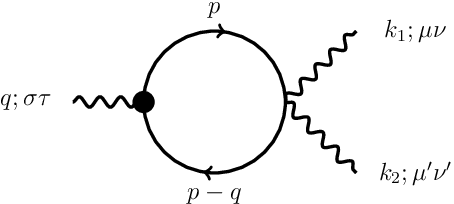}
\caption{\label{bubble} Bubble diagram with one $\overline{V}_{ffg}$ and one $V_{ffgg}$ corresponding to $i\Big\langle{\overline{T}^{(R,0)}}^{\sigma\tau}(x)S^{(2)}\Big\rangle_0$ term. }
\end{figure}

\begin{figure}[H]
\center
\includegraphics[width=0.5\textwidth]{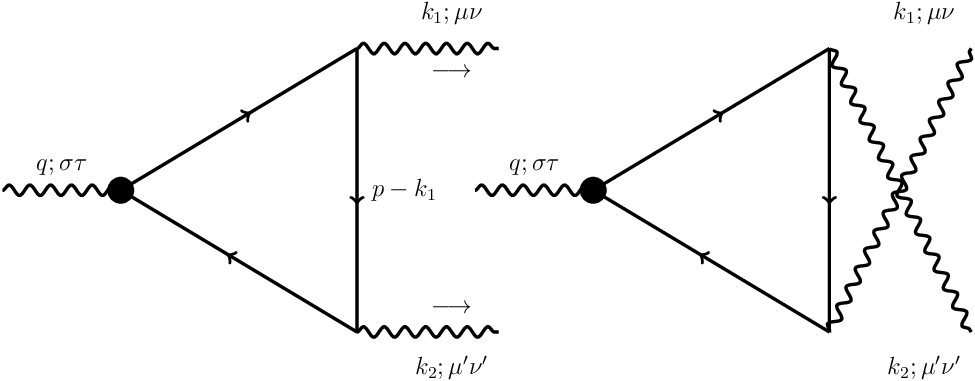}
\caption{\label{triangle} Triangle diagrams with one $\overline{V}_{ffg}$ and two $V_{fgg}$s corresponding to $-\frac{1}{2}\Big\langle{\overline{T}^{(R,0)}}^{\sigma}_{\sigma}(x)S^{(1)}S^{(1)}\Big\rangle_0$ term.  }
\end{figure}

Then we calculate the expectation value of ${\overline{T}^{(R)}}^{\sigma}_{\sigma}(x)$ at $\mathcal{O}(h^2)$ order. The expression is
\bea
&&\Big\langle{\overline{T}^{(R)}}^{\sigma}_{\sigma}(x)\Big\rangle\Big|_{\mathcal{O}(h^2)}=i\Big\langle{\overline{T}^{(R,1)}}^{\sigma}_{\sigma}(x)S^{(1)}\Big\rangle_0
+i\Big\langle{\overline{T}^{(R,0)}}^{\sigma}_{\sigma}(x)S^{(2)}\Big\rangle_0
-\frac{1}{2}\Big\langle{\overline{T}^{(R,0)}}^{\sigma}_{\sigma}(x)S^{(1)}S^{(1)}\Big\rangle_0 \nonumber\\
&&=\int d^4y \int \frac{d^4 q}{(2\pi)^4}{\rm{e}}^{-iq\cdot(x-y)}{T^{(1)}}^{\sigma}_{\sigma\mu\nu\mu'\nu'}(q,m)h^{\mu\nu}(y)h^{\mu'\nu'}(x)+\int d^4y \int d^4z \int \frac{d^4 k_1}{(2\pi)^4} \int \frac{d^4 k_2}{(2\pi)^4} \nonumber\\
&& \times{\rm{e}}^{-ik_1\cdot(x-y)}{\rm{e}}^{-ik_2\cdot(x-z)}\left({T^{(2)}}^{\sigma}_{\sigma\mu\nu\mu'\nu'}(k_1,k_2,m)+{\overline{T}^{(3)}}^{\sigma}_{\sigma\mu\nu\mu'\nu'}(k_1,k_2,m)\right)
h^{\mu\nu}(y)h^{\mu'\nu'}(z).
\eea
Here the Feynman diagrams of ${T^{(1)}}^{\sigma}_{\sigma\mu\nu\mu'\nu'}(q,m)$, ${T^{(2)}}^{\sigma}_{\sigma\mu\nu\mu'\nu'}(k_1,k_2,m)$ and ${\overline{T}^{(3)}}^{\sigma}_{\sigma\mu\nu\mu'\nu'}(k_1,k_2,m)$ are the figure \ref{oneloop3}, figure \ref{bubble} and figure \ref{triangle}, respectively.
\par

Using the Feynman rules (figure \ref{oneloop3}), the ${T^{(1)}}^{\sigma}_{\sigma\mu\nu\mu'\nu'}(q,m)$ is
\bea
\label{TToneloop3}
{T^{(1)}}^{\sigma}_{\sigma\mu\nu\mu'\nu'}(q,m)&=&\frac{1}{8}\int \frac{d^4 p}{(2\pi)^4}{\rm{tr}}\left[ \frac{1}{p\!\!\!/-m}\left((2p-q)_{\mu}\gamma_{\nu}+(\mu\leftrightarrow\nu)\right) \frac{1}{p\!\!\!/-q\!\!\!/-m}\right. \nonumber\\
&&\left.\times(-2\epsilon_{\mu'\nu' ad}q^a\gamma^d+\frac{m}{4}\eta_{\mu'\nu'})\frac{1+\gamma^5}{2}\right].
\eea
The result of (\ref{TToneloop3}) also has no Pontryagin term.

There other two diagrams (the bubble graph (figure \ref{bubble}) and the triangle graph (figure \ref{triangle})) have none zero contributions to the $\Big\langle{\overline{T}^{(R)}}^{\sigma}_{\sigma}(x)\Big\rangle^{odd}\Big|_{\mathcal{O}(h^2)}$. We set that the ingoing graviton has momentum $q$ and Lorentz labels $\sigma$, $\tau$ and two outgoing graviton are specified by $k_1$, $\mu$, $\nu$ and $k_2$, $\mu'$, $\nu'$, respectively. As the same as \cite{Bonora:2014qla}, we put the two external outgoing gravitons on-shell which means that the corresponding fields satisfy the EOM of gravity $R_{\mu\nu}=0$. We choose the de Donder gauge
\bea
\Gamma^{\lambda}_{\mu\nu}g^{\mu\nu}=0.
\eea
In momentum space this means that $k_1^2=k_2^2=0$.

The bubble diagram with two vertices is illustrated in figure \ref{bubble}. Its contribution to the trace anomaly comes from contracting the indices $\sigma$ and $\tau$ with $\eta^{\sigma\tau}$.
\par

The ${T^{(2)}}^{\sigma}_{\sigma\mu\nu\mu'\nu'}(k_1,k_2,m)$ is
\bea
&&{T^{(2)}}^{\sigma}_{\sigma\mu\nu\mu'\nu'}(k_1,k_2,m)= \frac{1}{512}\int \frac{d^4 p}{(2\pi)^4}t_{\mu\nu\mu'\nu'\lambda\rho}(k_2-k_1)^{\lambda} \nonumber\\
&&{\rm{tr}} \left(\frac{1}{p\!\!\!/-m}\gamma^{\rho}\gamma^5\frac{1}{p\!\!\!/-k\!\!\!/_1-k\!\!\!/_2-m} ( 2p\!\!\!/-k\!\!\!/_1-k\!\!\!/_2-2m) (1+\gamma^5)\right).\nonumber
\eea
We regulate the integral by introducing a set of Pauli-Villars fields \cite{Pauli:1949zm} with masses $M_i$ and compute
\bea \label{TTD}
{T^{(2),{\rm{reg}}}}^{\sigma}_{\sigma\mu\nu\mu'\nu'}(k_1,k_2,m)={T^{(2)}}^{\sigma}_{\sigma\mu\nu\mu'\nu'}(k_1,k_2,m)-\sum_{i}c_i {T^{(2)}}^{\sigma}_{\sigma\mu\nu\mu'\nu'}(k_1,k_2,M_i).
\eea
To remove the divergences, we impose the conditions
\bea
\label{condition}
\sum_{i}c_i=1, \quad \sum_{i}c_i M_i^2=m^2.
\eea
Taking the limit $M_i\rightarrow \infty$, the (\ref{TTD}) becomes

\bea
{T^{(2),{\rm{reg}}}}^{\sigma}_{\sigma\mu\nu\mu'\nu'}(k_1,k_2,m)=d(k_1\cdot k_2,m)k_1\cdot k_2t_{\mu\nu\mu'\nu'\lambda\rho}k_1^{\lambda}k_2^{\rho},
\eea
where the coefficient $d(k_1\cdot k_2,m)$ is
\bea \label{ddd1}
&&d(k_1\cdot k_2,m)=-\frac{i}{1536 \pi^2 (k_1\cdot k_2)^2}\left [-9m^2k_1\cdot k_2+(k_1\cdot k_2)^2-3m^2  \right.\nonumber\\
&& \left. \times\sqrt{k_1\cdot k_2(-2m^2+k_1\cdot k_2)}\log\left(\frac{m^2-k_1\cdot k_2+\sqrt{k_1\cdot k_2(-2m^2+k_1\cdot k_2)}}{m^2}\right)
  \right].
\eea

The triangle diagrams (figure \ref{triangle}) have three vertices, one $\overline{V}_{ffg}$ and two $V_{fgg}$s. The calculation of the full triangle diagrams ${\overline{T}^{(3)}}^{\sigma}_{\sigma\mu\nu\mu'\nu'}(k_1,k_2,m)$ are complicated. The evaluation of the Pontryagin term which is contained in the $\gamma^5$ sector is easier that the full triangle diagrams.

\par

The $\gamma^5$ sector of full triangle diagrams is
\bea
\label{TT}
&{T^{(3)}}^{\sigma}_{\sigma\mu\nu\mu'\nu'}(k_1,k_2,m)=\frac{i}{256}\int \frac{d^4 p}{(2\pi)^4}{\rm{tr}}\left[ \frac{1}{p\!\!\!/-m}\left((2p-k_1)_{\mu}\gamma_{\nu}+(\mu\leftrightarrow\nu)\right) \frac{1}{p\!\!\!/-k\!\!\!/_1-m}\right. \nonumber\\
&\left.\times \left((2p-2k_1-k_2)_{\mu'}\gamma_{\nu'}+(\mu'\leftrightarrow\nu')\right) \frac{1}{p\!\!\!/-k\!\!\!/_1-k\!\!\!/_2-m}(2p\!\!\!/-k\!\!\!/_1-k\!\!\!/_2-2m)\frac{\gamma^5}{2}\right] \nonumber\\
&+(k_1,\mu,\nu\leftrightarrow k_2,\mu',\nu').
\eea
As the same as the bubble diagram, we regulate the integral by introducing a set of Pauli-Villars fields \cite{Pauli:1949zm} with masses $M_i$ and compute
\bea \label{TT1}
{T^{(3),{\rm{reg}}}}^{\sigma}_{\sigma\mu\nu\mu'\nu'}(k_1,k_2,m)={T^{(3)}}^{\sigma}_{\sigma\mu\nu\mu'\nu'}(k_1,k_2,m)-\sum_{i}c_i {T^{(3)}}^{\sigma}_{\sigma\mu\nu\mu'\nu'}(k_1,k_2,M_i).
\eea
Then we impose the conditions (\ref{condition}) and
take the limit $M_i\rightarrow \infty$, the (\ref{TT1}) becomes
\bea
&&{T^{(3),{\rm{reg}}}}^{\sigma}_{\sigma\mu\nu\mu'\nu'}(k_1,k_2,m)=\nonumber \\
&&\left(a(k_1\cdot k_2,m)t_{\mu\nu\mu'\nu'\kappa\lambda}^{(21)}
+b(k_1\cdot k_2,m) k_1\cdot k_2t_{\mu\nu\mu'\nu'\kappa\lambda}
+c(k_1\cdot k_2,m) t_{\mu\nu\mu'\nu'\kappa\lambda}^{(12)} \right) k_1^{\kappa}k_2^{\lambda}.\nonumber
\eea
Where the coefficients $a(k_1\cdot k_2,m)$, $b(k_1\cdot k_2,m)$ and $c(k_1\cdot k_2,m)$ are
\bea \label{abc}
&& a(k_1\cdot k_2,m)=i\frac{6m^2 k_1\cdot k_2 +(k_1\cdot k_2)^2+3m^4\log^2\left(\frac{m^2-k_1\cdot k_2+\sqrt{k_1\cdot k_2(-2m^2+k_1\cdot k_2)}}{m^2}\right)}{3072 \pi^2 (k_1\cdot k_2)^2},\nonumber\\
&& b(k_1\cdot k_2,m)=i\frac{1}{3072 \pi^2 (k_1\cdot k_2)^2}\left [(k_1\cdot k_2)^2 \right.\nonumber\\
&& \left. -6m^2 \sqrt{k_1\cdot k_2(-2m^2+k_1\cdot k_2)}\log\left(\frac{m^2-k_1\cdot k_2+\sqrt{k_1\cdot k_2(-2m^2+k_1\cdot k_2)}}{m^2}\right)\right. \nonumber \\
&& \left. -3m^4\log^2\left(\frac{m^2-k_1\cdot k_2+\sqrt{k_1\cdot k_2(-2m^2+k_1\cdot k_2)}}{m^2}\right)-24m^2k_1\cdot k_2 \right], \nonumber\\
&&c(k_1\cdot k_2,m)=i\frac{m^2}{2048 \pi^2 (k_1\cdot k_2)^2}\left [2\log\left(\frac{m^2-k_1\cdot k_2+\sqrt{k_1\cdot k_2(-2m^2+k_1\cdot k_2)}}{m^2}\right) \right.\nonumber\\
&& \left.\left(4\sqrt{k_1\cdot k_2(-2m^2+k_1\cdot k_2)}
+m^2\log\left(\frac{m^2-k_1\cdot k_2+\sqrt{k_1\cdot k_2(-2m^2+k_1\cdot k_2)}}{m^2}\right)\right)\right. \nonumber \\
&&\left.+k_1\cdot k_2\left(20+\log^2\left(\frac{m^2-k_1\cdot k_2+\sqrt{k_1\cdot k_2(-2m^2+k_1\cdot k_2)}}{m^2}\right)\right)\right ]
\eea
and the quantity $t_{\mu\nu\mu'\nu'\kappa\lambda}^{(21)}$ is defined to be
\bea
t_{\mu\nu\mu'\nu'\kappa\lambda}^{(21)}=k_{2\mu}k_{1\mu'}\epsilon_{\nu\nu'\kappa\lambda}+k_{2\nu}k_{1\nu'}\epsilon_{\mu\mu'\kappa\lambda}
+k_{2\mu}k_{1\nu'}\epsilon_{\nu\mu'\kappa\lambda}+k_{2\nu}k_{1\mu'}\epsilon_{\mu\nu'\kappa\lambda}.
\eea
Putting ${T^{(2),{\rm{reg}}}}^{\sigma}_{\sigma\mu\nu\mu'\nu'}(k_1,k_2,m)$ and ${T^{(3),{\rm{reg}}}}^{\sigma}_{\sigma\mu\nu\mu'\nu'}(k_1,k_2,m)$ together, we get
\bea
&&{T^{(2),{\rm{reg}}}}^{\sigma}_{\sigma\mu\nu\mu'\nu'}(k_1,k_2,m)+{T^{(3),{\rm{reg}}}}^{\sigma}_{\sigma\mu\nu\mu'\nu'}(k_1,k_2,m)=
\left(a(k_1\cdot k_2,m)t_{\mu\nu\mu'\nu'\kappa\lambda}^{(21)} \right. \nonumber \\
&&\left.+\left(b(k_1\cdot k_2,m)+ d(k_1\cdot k_2,m)\right)k_1\cdot k_2t_{\mu\nu\mu'\nu'\kappa\lambda}+c(k_1\cdot k_2,m) t_{\mu\nu\mu'\nu'\kappa\lambda}^{(12)} \right) k_1^{\kappa}k_2^{\lambda}.
\eea
From (\ref{ddd1}) and (\ref{abc}), we have the relation
\bea
b(k_1\cdot k_2,m)+ d(k_1\cdot k_2,m)=-a(k_1\cdot k_2,m).
\eea
This indicates that the ${T^{(2),{\rm{reg}}}}^{\sigma}_{\sigma\mu\nu\mu'\nu'}(k_1,k_2,m)+{T^{(3),{\rm{reg}}}}^{\sigma}_{\sigma\mu\nu\mu'\nu'}(k_1,k_2,m)$ satisfied the diffeomorphisms Ward identities
\bea
\label{ward}
&&k_1^{\mu}\left({T^{(2),{\rm{reg}}}}^{\sigma}_{\sigma\mu\nu\mu'\nu'}(k_1,k_2,m)+{T^{(3),
{\rm{reg}}}}^{\sigma}_{\sigma\mu\nu\mu'\nu'}(k_1,k_2,m)\right)=0, \nonumber \\
&&k_2^{\mu'}\left({T^{(2),{\rm{reg}}}}^{\sigma}_{\sigma\mu\nu\mu'\nu'}(k_1,k_2,m)+{T^{(3),
{\rm{reg}}}}^{\sigma}_{\sigma\mu\nu\mu'\nu'}(k_1,k_2,m)\right)= 0.
\eea
Taking massless limit ($m\rightarrow 0$), the ${T^{(2),{\rm{reg}}}}^{\sigma}_{\sigma\mu\nu\mu'\nu'}(k_1,k_2,m)+{T^{(3),{\rm{reg}}}}^{\sigma}_{\sigma\mu\nu\mu'\nu'}(k_1,k_2,m)$ becomes
\bea
&&{T^{(2),{\rm{reg}}}}^{\sigma}_{\sigma\mu\nu\mu'\nu'}(k_1,k_2,m)+{T^{(3),{\rm{reg}}}}^{\sigma}_{\sigma\mu\nu\mu'\nu'}(k_1,k_2,m)=\nonumber \\
&&-\frac{i}{3072 \pi^2 }\left(k_1\cdot k_2t_{\mu\nu\mu'\nu'\kappa\lambda}-t_{\mu\nu\mu'\nu'\kappa\lambda}^{(21)}
\right) k_1^{\kappa}k_2^{\lambda}.\nonumber
\eea
Putting all contributions together, we obtain
\bea
\label{PontryaginPontryagin1}
&&\Big\langle{\overline{T}^{(R)}}^{\sigma}_{\sigma}(x)\Big\rangle^{odd}=-\frac{i}{3072 \pi^2 }\int d^4y \int d^4z \int \frac{d^4 k_1}{(2\pi)^4} \int \frac{d^4 k_2}{(2\pi)^4}{\rm{e}}^{-ik_1\cdot(x-y)}{\rm{e}}^{-ik_2\cdot(x-z)} \nonumber\\
&&\times\left(\left(k_1\cdot k_2t_{\mu\nu\mu'\nu'\kappa\lambda}-t_{\mu\nu\mu'\nu'\kappa\lambda}^{(21)}\right) k_1^{\kappa}k_2^{\lambda}\right)h^{\mu\nu}(y)h^{\mu'\nu'}(z)
+\ldots \nonumber\\
&&=-\frac{i}{768\pi^2}\epsilon_{\nu\nu'\kappa\lambda}\left(\partial_{\mu'}\partial^{\kappa}h^{\mu\nu}\partial_{\mu}\partial^{\lambda}h^{\mu'\nu'}
-\partial^{\kappa}\partial_{\rho}h^{\mu\nu}\partial^{\lambda}\partial^{\rho}h^{\nu'}_{\mu}\right)+\ldots.
\eea
The Pontryagin density has the following approximation in the metric fluctuation $h_{\mu\nu}$
\bea
\label{PontryaginPontryagin2}
\epsilon_{\nu\nu'\kappa\lambda}R^{\nu\nu'}_{\ \ \rho\mu}R^{\rho\mu\kappa\lambda}=-2\epsilon_{\nu\nu'\kappa\lambda}
\left(\partial_{\mu'}\partial^{\kappa}h^{\mu\nu}\partial_{\mu}\partial^{\lambda}h^{\mu'\nu'}
-\partial^{\kappa}\partial_{\rho}h^{\mu\nu}\partial^{\lambda}\partial^{\rho}h^{\nu'}_{\mu}\right)+\ldots.
\eea
Comparing the expression (\ref{PontryaginPontryagin2}) with formula (\ref{PontryaginPontryagin1}), we finally get the result
\bea
\label{anomalyanomaly}
\Big\langle{\overline{T}^{(R)}}^{\sigma}_{\sigma}(x)\Big\rangle^{odd}=\frac{i}{1536\pi^2}\epsilon_{\nu\nu'\kappa\lambda}R^{\nu\nu'}_{\ \ \rho\mu}R^{\rho\mu\kappa\lambda}.
\eea

The
coefficient of the parity-odd term is purely imaginary signaling a violation of unitarity. To solve this problem, the authors \cite{Bonora:2014qla} had an argument that the neutrino can have mass. We have another idea on this problem. In our previous work \cite{Liu:2018dfm}, we connected the muli-valued function with bound state. Basing on this idea, we added the bound state contribution to get an anomaly free theory \cite{Liu:2021pmt}. We can use this method to obtain the zero Pontryagin density by taking consider the bound state contribution.
We note that the complex function $\log(x)$ is multi-valued, that is
\bea
\log(x)={\rm{Log}}(x)+2\pi n i,\quad n\in Z.
\eea
Where the argument of function ${\rm{Log}}(x)$ is $\arg({\rm{Log}}(x))\in (-\pi,\pi]$. Putting this new expression of $\log(x)$ into the (\ref{abc}), the function $a(k_1\cdot k_2,m)$
becomes
\bea
\label{abc1}
&& a(k_1\cdot k_2,m)=i\frac{6m^2 k_1\cdot k_2 +(k_1\cdot k_2)^2+3m^4\left[\log\left(\frac{m^2-k_1\cdot k_2+\sqrt{k_1\cdot k_2(-2m^2+k_1\cdot k_2)}}{m^2}\right)+2\pi n i\right]^2}{3072 \pi^2 (k_1\cdot k_2)^2}.\nonumber
\eea
We find that the $a(k_1\cdot k_2,m)$ becomes zero in the massless limit as long as
\bea
q^2=(k_1+k_2)^2=2k_1\cdot k_2=4\sqrt{3}n\pi m^2, \quad n\in \{0,N\}.
\eea
So the bubble (figure \ref{bubble}) and triangle diagrams (figure \ref{triangle}) represent a physical process that is a neutral bound state decay into two gravitons. This is the same as the pion decay $\pi^0\rightarrow \gamma\gamma$.

\section{Fujikawa's method}
Following the works of \cite{Diaz:1989nx,Bastianelli:2016nuf}, we use the method of Fujikawa \cite{Fujikawa:1979ay,Fujikawa:1980vr} to evaluate anomalies. We consider Lagrangian of a Dirac fermion $\psi$ in a curved spacetime, which is
\bea
\label{Lagrangian1}
\overline{\mathcal{L}}=\sqrt{|g|}\left[i\overline{\psi}\gamma^{\mu}\nabla_{\mu}P_R\psi-m\overline{\psi}P_R\psi\right]=
\sqrt{|g|}\left[i\overline{\psi}\nabla\!\!\!\!/P_R\psi-m\overline{\psi}P_R\psi\right],
\eea
where the $\nabla\!\!\!\!/$ is $\nabla\!\!\!\!/=\gamma^{\mu}\nabla_{\mu}=e^{\mu}_a\gamma^a \nabla_{\mu}$. The Lagrangian (\ref{Lagrangian1}) which we will consider is different with
the one in the paper \cite{Bastianelli:2016nuf}.
In massless limit $m=0$, the Lagrangian (\ref{Lagrangian1}) has the form of a Weyl fermion
\bea
\widehat{\mathcal{L}}=\overline{\mathcal{L}}\Big|_{m=0}=\sqrt{|g|}i\overline{\psi}\nabla\!\!\!\!/P_R\psi=\sqrt{|g|}i\overline{\psi_R}\nabla\!\!\!\!/\psi_R.
\eea
The corresponding action $\widehat{S}$
\bea
\label{Lagrangian2}
\widehat{S}=\int d^4x
\sqrt{|g|}i\overline{\psi}\nabla\!\!\!\!/P_R\psi
\eea
 is invariant under conformal transformation (\ref{conformal1}). The other fields transform accordingly by the rules
\bea
\label{conformal2}
\psi(x)&\rightarrow& \psi'(x)={\rm{e}}^{-\frac{3}{2} \sigma(x)} \psi(x),\nonumber \\
\overline{\psi}(x)&\rightarrow& \overline{\psi}'(x)={\rm{e}}^{-\frac{3}{2} \sigma(x)} \overline{\psi}(x),\nonumber \\
{\rm{e}}^a_{\mu} &\rightarrow& {\rm{e}}'^a_{\mu}={\rm{e}}^{\sigma(x)} {\rm{e}}^a_{\mu}.
\eea

To get a symmetric form of Lagrangian, we put the dynamical variables into a column vector $\phi$ as
\bea
\phi=\left(
  \begin{array}{c}
    \psi  \\
    C^{-1}\overline{\psi}^T  \\
  \end{array}
\right)=
\left(
  \begin{array}{c}
    \psi  \\
    \psi_c  \\
  \end{array}
\right).
\eea
The $\psi_c$ is the charge conjugated field
\bea
\psi_c= C^{-1}\overline{\psi}^T.
\eea
Then the Lagrangian (\ref{Lagrangian1}) has the following symmetric form
\bea
\overline{\mathcal{L}}=-\frac{1}{2}\phi^T T\mathcal{O}P\phi+\frac{1}{2} m \phi^T T \widetilde{P}\phi.
\eea
Here the matrix $T$, $\mathcal{O}$, $P$ and $\widetilde{P}$ are defined as
\bea
\label{TOP}
&&T=\left(
  \begin{array}{cc}
    0 &\ \ \sqrt{|g|}C^T \\
    \sqrt{|g|}C &\ \ 0 \\
  \end{array}
\right),
\quad
\mathcal{O}=\left(
  \begin{array}{cc}
    i\nabla\!\!\!\!/ &\ \ 0 \\
    0 &\ \ i\nabla\!\!\!\!/ \\
  \end{array}
\right),\nonumber
\\
&&P=\left(
  \begin{array}{cc}
    P_R &\ \ 0 \\
    0 &\ \ P_L \\
  \end{array}
\right),
\quad
\widetilde{P}=\left(
  \begin{array}{cc}
    P_R &\ \ 0 \\
    0 &\ \ P_R \\
  \end{array}
\right).
\eea
To regulate the divergences, we introduce the Pauli-Villars field $\theta$ with mass $M$. The Lagrangian of Pauli-Villars field is
\bea
\label{Lagrangian3}
\overline{\mathcal{L}}_{{\rm{PV}}}=
\sqrt{|g|}\left[i\overline{\theta}\nabla\!\!\!\!/P_R\theta-M\overline{\theta}P_R\theta\right].
\eea
Collecting the PV fields by
\bea
\chi=
\left(
  \begin{array}{c}
    \theta  \\
    \theta_c  \\
  \end{array}
\right),
\eea
the Lagrangian (\ref{Lagrangian3}) is rewritten in the form
\bea
\label{Lagrangian4}
\overline{\mathcal{L}}_{{\rm{PV}}}=-\frac{1}{2}\chi^T T\mathcal{O}P\chi+\frac{1}{2} M \chi^T T \widetilde{P}\chi.
\eea
The matrix $T$, $\mathcal{O}$, $P$ and $\widetilde{P}$ are the same as (\ref{TOP}).

For simplifying our discussion, we denote $\delta\phi=K\phi$ and $\delta\chi=K\chi$ by the infinitesimal Weyl transformation on $\phi$ and $\chi$, respectively. From the conformal transformation
 (\ref{conformal2}), we find that $K=-\frac{3}{2} \sigma(x)$.

We now calculate the Pontryagin term. The regulated action has the form
\bea
\widetilde{S}=\int d^4x \left(\widehat{\mathcal{L}}+\overline{\mathcal{L}}_{{\rm{PV}}}\right).
\eea
The quantum theory is defined by the path integral
\bea
{\rm{e}}^{i\widetilde{W}}=Z=\int \mathcal{D}\phi\mathcal{D}\chi {\rm{e}}^{i\widetilde{S}}.
\eea
From this, we obtain
\bea
\delta W=-i\frac{\delta Z}{Z}=\frac{\int \mathcal{D}\phi\mathcal{D}\chi \delta \widetilde{S}{\rm{e}}^{i\widetilde{S}}}{\int \mathcal{D}\phi\mathcal{D}\chi {\rm{e}}^{i\widetilde{S}}}
=\langle\delta \widetilde{S}\rangle.
\eea
According to the definition (\ref{delta}), the trace anomaly is obtained by the computing the quantity
\bea
\mathcal{A}=\int d^4x\ \sigma(x) \sqrt{|g|}\langle T^{\mu}_{\mu}\rangle=-\langle\delta \widetilde{S}\rangle=i\frac{\delta Z}{Z}.
\eea
Where the $\mathcal{A}$ is connected with trace anomaly $A$ (\ref{traceanomaly}) by the relation
\bea
\mathcal{A}=\int d^4x\ \sigma(x) \sqrt{|g|}A.
\eea
The $\delta Z$ is calculated directly
\bea
\delta Z&=&\int \mathcal{D}\phi\mathcal{D}\chi {\rm{e}}^{i\widetilde{S}(\phi',\chi')}-\int \mathcal{D}\phi\mathcal{D}\chi {\rm{e}}^{i\widetilde{S}(\phi,\chi)} \nonumber \\
&=&\int\mathcal{D}\phi\mathcal{D}\chi {\rm{e}}^{i\widehat{S}(\phi)+i\int d^4x \left[\frac{1}{2}M\chi^T(TK+K^T T+\delta T)\widetilde{P}\chi\right]}-\int \mathcal{D}\phi\mathcal{D}\chi {\rm{e}}^{i\widetilde{S}(\phi,\chi)}\nonumber \\
&=&\int\mathcal{D}\phi \left({\rm{det}}\left[M(TK+K^T T+\delta T)\widetilde{P}\right]\right)^{-\frac{1}{2}}{\rm{e}}^{i\widehat{S}(\phi)}-\int \mathcal{D}\phi\mathcal{D}\chi {\rm{e}}^{i\widetilde{S}(\phi,\chi)}.
\eea
Here the jacobian of Pauli-Villars fields cancels the jacobian of the original fields $\phi$. The $P$ and $\widetilde{P}$ are not invertible, so we can not write the resulting determinant as a product of the determinant $- T\mathcal{O}P+M T\widetilde{P}$ and others. As the authors \cite{Bonora:2019dyv} claimed that the lack of an inverse for the chiral Weyl
kinetic term has drastic consequences. So we factor the determinant as the following
\bea
&&{\rm{det}}\left[M(TK+K^T T+\delta T)\widetilde{P}\right]= \nonumber\\
&&{\rm{det}}\left[M(TK+K^T T+\delta T)\widetilde{P}(- T\mathcal{O}+M T )^{-1}(- T\mathcal{O}+M T )\right].
\eea
Then the $\delta Z$ becomes
\bea
\label{deltaz1}
\delta Z=\int\mathcal{D}\phi \mathcal{D}\chi \left({\rm{det}}\left[M(TK+K^T T+\delta T)\widetilde{P}(- T\mathcal{O}+M T )^{-1}\right]\right)^{-\frac{1}{2}}{\rm{e}}^{i\widetilde{S}'(\phi,\chi)}-Z.
\eea
Where the new action $\widetilde{S}'(\phi,\chi)$ is defined by
\bea
\widetilde{S}'(\phi,\chi)=\int d^4x \left(\widehat{\mathcal{L}}-\frac{1}{2}\chi^T T\mathcal{O}\chi+\frac{1}{2} M \chi^T T \chi\right).
\eea

The determinant in expression (\ref{deltaz1}) may be written as
\bea
{\rm{det}}\left[M(TK+K^T T+\delta T)\widetilde{P}(- T\mathcal{O}+M T )^{-1}\right]= \nonumber\\
1+{\rm{Tr}}\left[M(TK+K^T T+\delta T)\widetilde{P}(- T\mathcal{O}+M T )^{-1}\right].
\eea
Then the $\mathcal{A}$ can be expressed as
\bea
\label{Trace}
\mathcal{A}&=&-\lim_{M\rightarrow \infty}\frac{i}{2}{\rm{Tr}}\left[M(TK+K^T T+\delta T)\widetilde{P}(- T\mathcal{O}+M T )^{-1}\right] \nonumber\\
&=&-i\lim_{M\rightarrow \infty}{\rm{Tr}}\left[\left(K+\frac{1}{2}T^{-1}\delta T \right)\widetilde{P}\left(1-\frac{\mathcal{O}}{M}\right)^{-1}
\right].
\eea
Inserting the identity $1=\left(1+\frac{\mathcal{O}}{M}\right)\left(1+\frac{\mathcal{O}}{M}\right)^{-1}$ into (\ref{Trace}), the
$\mathcal{A}$ becomes
\bea
\label{Trace1}
\mathcal{A}=
-i\lim_{M\rightarrow \infty}{\rm{Tr}}\left[\left(K\widetilde{P}+\frac{1}{2}T^{-1}\delta T\widetilde{P}
 +K\frac{\widetilde{P}\mathcal{O}}{M}+\frac{1}{2}T^{-1}\delta T\frac{\widetilde{P} \mathcal{O}}{M}\right)\left(1-\left(\frac{\mathcal{O}}{M}\right)^2\right)^{-1}
\right].\nonumber
\eea
Using the background invariance of the kinetic term
\bea
\label{definition}
\phi^T\left(T\mathcal{O}K+\frac{1}{2}\delta (T\mathcal{O})\right)\phi=0,
\eea
the $\mathcal{A}$ has the form
\bea
\label{Trace2}
\mathcal{A}&=&-i\lim_{M\rightarrow \infty}{\rm{Tr}}\left[\left(K\widetilde{P}+\frac{1}{2}T^{-1}\delta T \widetilde{P}-\frac{1}{2}\frac{\widetilde{P}\mathcal{\delta O}}{M}\right)\left(1-\left(\frac{\mathcal{O}}{M}\right)^2\right)^{-1}
\right] \nonumber\\
&=&-i\lim_{M\rightarrow \infty}{\rm{Tr}}\left[\left(K\widetilde{P}+\frac{1}{2}T^{-1}\delta T\widetilde{P} -\frac{1}{2}\frac{\widetilde{P} \mathcal{\delta O}}{M}\right){\rm{e}}^
{\frac{\left(\mathcal{O}\right)^2}{M^2}}
\right].
\eea
From the equation (\ref{definition}), we obtain
\bea
\label{delta33}
\delta \mathcal{O}=i\left(
  \begin{array}{cc}
   -\sigma \nabla\!\!\!\!/+3( \partial\!\!\!/ \sigma) &\ \  0 \\
    0 &\ \ -\sigma \nabla\!\!\!\!/+3( \partial\!\!\!/ \sigma) \\
  \end{array}
\right).
\eea
Substituting (\ref{delta33}) into the expression (\ref{Trace2}), we find
\bea
\label{Trace3}
\mathcal{A}=-i\lim_{M\rightarrow \infty}{\rm{Tr}}\left[\widetilde{P}\widetilde{\mathcal{Q}}{\rm{e}}^
{\frac{\left(\mathcal{O}\right)^2}{M^2}}
\right],
\eea
where the matrix $\widetilde{\mathcal{Q}}$ is
\bea
\widetilde{\mathcal{Q}}=\left(
  \begin{array}{cc}
    \frac{\sigma}{2}+\frac{i}{2M}\left(\sigma \nabla\!\!\!\!/-3( \partial\!\!\!/ \sigma)\right) &\ \ 0  \\
    0 &\ \ \frac{\sigma}{2}+\frac{i}{2M}\left(\sigma \nabla\!\!\!\!/-3( \partial\!\!\!/ \sigma)\right) \\
  \end{array}
\right).
\eea
We only consider the Pontryagin term which is obtained from the $\gamma^5$ sector in (\ref{Trace3}). As a non-vanishing Dirac trace with the $\gamma^5$ requires at least four $\gamma$-matrices. The expression (\ref{Trace3}) can be simplified to

\bea
\mathcal{A}=-i\lim_{M\rightarrow \infty}{\rm{Tr}}\left[\frac{\sigma}{2}\widetilde{P}
{\rm{e}}^{\frac{(i\nabla\!\!\!\!/)^2}{M^2}} \right]
=-i\lim_{M\rightarrow \infty}{\rm{Tr}}\left[\sigma P_R
{\rm{e}}^{\frac{(i\nabla\!\!\!\!/)^2}{M^2}} \right]_4.
\eea
Where the ${\rm{Tr}}\left[\cdots\right]_4$ represents trace on the four dimensional Dirac matrices. The $\mathcal{A}^{odd}$ which is associated with parity-odd anomaly term is
\bea
\label{oddodd}
\mathcal{A}^{odd}=-\frac{i}{2}\lim_{M\rightarrow \infty}{\rm{Tr}}\left[\gamma^5 \sigma
{\rm{e}}^{\frac{-(\nabla\!\!\!\!/)^2}{M^2}} \right]_4=-\frac{i}{2}a_4(\gamma^5 \sigma,D).
\eea
Where the $D$ is $D=(\nabla\!\!\!\!/)^2$. We have used the heat kernel method (see appendix B or review paper \cite{Vassilevich:2003xt}) to get the expression (\ref{oddodd}). From (\ref{a4}) in appendix B, we obtain

\bea
\mathcal{A}^{odd}&=&-\frac{i}{2}\frac{1}{(4\pi)^2}\frac{1}{12\times 16}(-i)\int d^4x \sigma \sqrt{|g|}R_{\sigma\rho\mu\nu}R^{\mu\nu}_{\ \ ij}{\rm{Tr}}\left(\gamma^5
\gamma^{\sigma}\gamma^{\rho}\gamma^{i}\gamma^{j}\right) \nonumber \\
&=&\frac{i}{1536\pi^2}\int d^4x \sigma \sqrt{|g|}R_{\sigma\rho\mu\nu}R^{\mu\nu}_{\ \ ij}\epsilon^{\sigma\rho ij}.
\eea
Where the extra $(-i)$ comes from Euclidean space back to Minkowski space-time. The parity-odd term of trace anomaly is related to the Kimura-Delbourgo-Salam anomaly \cite{Kimura:1969iwz,Delbourgo:1972xb} which is the anomalous divergence of the
axial current $\langle i\overline{\psi}\gamma^{\mu}\gamma^5\psi\rangle$ in a gravitational field, that is

\bea
A^{odd}=\frac{1}{4}\nabla_{\mu}\langle i\overline{\psi}\gamma^{\mu}\gamma^5\psi\rangle=\frac{i}{1536\pi^2}R_{\sigma\rho\mu\nu}R^{\mu\nu}_{\ \ ij}\epsilon^{ij\sigma\rho }.
\eea
Then the parity-odd term of trace anomaly obtained by the Fujikawa's method is the same as the one by Feynman diagram method (\ref{anomalyanomaly}). The final result is the same as the
work \cite{Bonora:2014qla}.

\section{Conclusions and Discussions }
In this paper, we have studied the Pontryagin trace anomaly for chiral fermions in a general
curved background using Pauli-Villars regularization. To introduce massive PV fields, we have utilized a Dirac mass term. After taking the massless limit, we recovered the chiral field theory by inserting the chiral projector in proper place of Dirac field Lagrangian. We used both Feynman diagram method and Fujikawa's method to calculate the parity-odd contribution. Our result indicates that the trace anomaly of energy-momentum tensor for chiral fermions has Pontryagin term
\bea
P=\frac{i}{1536\pi^2}\epsilon_{\nu\sigma\kappa\lambda}R^{\nu\sigma}_{\ \ \rho\mu}R^{\rho\mu\kappa\lambda}.
\eea
This agrees with the work of Bonora et al \cite{Bonora:2014qla}.

In our work, we have used the Pauli-Villars regularization instead of dimensional regularization. The reason is that how to treat the $\gamma^5$ matrix is still open problem in dimensional regularization. In our next work, we will apply the other prescriptions of dimensional regularization to investigate the trace anomaly of chiral fermions in a
curved background.



\section*{Acknowledgements}
This work is supported by Chinese Universities Scientific Fund Grant No. 2452018158. We would like to thank Dr. Wei He, Youwei Li and Suzhi Wu for helpful discussions.
\appendix
\section{Conventions and notations}

We use the flat metric $\eta_{\mu\nu}={\rm{diag}}(+1,-1,-1,-1)$ and work in units with $\hbar=c=1$. The chiral matrix $\gamma^5$ is defined by
\bea
\gamma^5=i\gamma^0\gamma^1\gamma^2\gamma^3.
\eea
It allows to define the left and right chiral projectors
\bea
P_{L}=\frac{1-\gamma^5}{2}, \quad\ P_{R}=\frac{1+\gamma^5}{2}.
\eea
A Dirac spinor $\psi$ can be split into two Weyl spinors
\bea
\psi=\psi_{L}+\psi_{R}=P_{L}\psi+P_{R}\psi.
\eea
The charge conjugation matrix $C$ is the matrix that satisfy
\bea
C\gamma^{\mu} C^{-1}=-(\gamma^{\mu})^{T}.
\eea
We take $C=i\gamma^0\gamma^2$ which has the properties
\bea
C=-C^T=-C^{-1}=-C^{\dag}=C^{\ast}.
\eea

 \section{The heat kernel method}
In this appendix, we collect some definitions and useful formulae from the review paper \cite{Vassilevich:2003xt}. Let $M$ be a smooth compact Riemannian manifold of dimension $n$. The $g_{\mu\nu}$ and $\omega_{\mu}$ are metric tensor and spin connection of $M$ respectively.
The field strength of the connection $\omega$ is
\bea
\Omega_{\mu\nu}=\partial_{\mu}\omega_{\nu}-\partial_{\nu}\omega_{\mu}+\omega_{\mu}\omega_{\nu}-\omega_{\nu}\omega_{\mu}.
\eea
The Riemann curvature tensor is
\bea
R^{\mu}_{\ \nu\rho\sigma}=\partial_{\sigma}\Gamma^{\mu}_{\nu\rho}-\partial_{\rho}\Gamma^{\mu}_{\nu\sigma}+\Gamma^{\lambda}_{\nu\rho}
\Gamma^{\mu}_{\lambda\sigma}-\Gamma^{\lambda}_{\nu\sigma}\Gamma^{\mu}_{\lambda\rho}.
\eea
 Let $D$ be self-adjoint operator and $f$ be an
auxiliary smooth function on $M$. There is an asymptotic
expansion as $t\rightarrow 0$
\bea
{\rm{Tr}}_{L^2}\left(f{\rm{e}}^{-tD}\right)\cong \sum_{k\geq 0}t^{\frac{(k-n)}{2}}a_k(f,D).
\eea

The
leading heat kernel coefficients $a_4(f,D)$ is known as \cite{DeWitt:1964mxt,McKean:1967xf}
\bea
\label{a4}
a_4(f,D)&=&\frac{1}{360}\frac{1}{(4\pi)^{\frac{n}{2}}}\int_M d^nx \sqrt{g}{\rm{tr}}_V\{f(60E_{;k}^{\ \ k}+60RE+180E^2 \nonumber\\
&&+12R_{;k}^{\ \ k}
+5R^2-2R_{ij}R^{ij}+2R_{ijkl}R^{ijkl}+30\Omega_{ij}\Omega^{ij})\}.
\eea
Where the $;$ denotes multiple covariant differentiation with respect to the Levi-Civita connection of $M$. The $R_{\mu\nu}:=
R^{\sigma}_{\ \mu\nu\sigma}$ is the Ricci tensor and $R:=R^{\mu}_{\mu}$ is the scalar curvature.
Let $\nabla\!\!\!\!/$ be the standard Dirac operator in curved space, that is
\bea
\label{nabla}
\nabla\!\!\!\!/=\gamma^{\mu}(\partial_{\mu}+\frac{1}{2}\omega_{\mu}),
\eea
then the $E$ and $\Omega_{\mu\nu}$ associated with $\nabla\!\!\!\!/$ are
\bea
E=-\frac{1}{4}R, \quad \Omega_{\mu\nu}=-\frac{1}{4}\gamma^{\sigma}\gamma^{\rho}R_{\sigma\rho\mu\nu}.
\eea


\end{document}